\documentclass[%
 reprint,
 amsmath,amssymb,
 aps,
pra,
]{revtex4-2}

\usepackage{graphicx}
\usepackage{dcolumn}
\usepackage{bm}
\usepackage{hyperref}
\usepackage{siunitx}
\usepackage{physics}

\begin{document}

\preprint{APS/123-QED}

\title{Runaway Electrons in Stellarators:\\Unlikely or Unavoidable?}

\author{Ida Ekmark}
\email{ida.ekmark@chalmers.se}
\affiliation{%
 Department of Physics and Astronomy, Chalmers University of Technology,\\ G\"{o}teborg, SE-41296, Sweden
}

\author{Matt Landreman}
\affiliation{
 Institute for Research in Electronics and Applied Physics, University of Maryland, College Park, MD 20742, USA
}

\date{November 25, 2026}

\begin{abstract}
Generation of relativistic runaway electrons has historically not been considered a possible problem in stellarators, but this may not hold in reactor-scale stellarators despite the lack of an externally driven plasma current. 
The magnitude of the plasma current governs the exponential generation of runaways, and even if there is no externally driven plasma current, the bootstrap current could be considerable in reactor-relevant stellarators. 
In this paper, we present a pilot study on the generation of runaway electrons in stellarator temperature collapse scenarios. 
To reliably study runaway electrons in stellarators, we implemented a stellarator plasma model in the runaway electron simulation tool \textsc{Dream}. 
The model is used to explore when runaway electrons can be generated with regard to combinations of initial plasma current, temperature decay time scale, and post-decay temperature. 
Special consideration is given to runaway generation through avalanche multiplication in stellarators, and how it compares to tokamaks. 
We find that significant runaway electron generation is possible also in stellarators, and demonstrate under which conditions it could be a concern. 
However, our findings support the conception that runaway electrons will be less of a concern in reactor-scale stellarators compared to tokamaks. 
\end{abstract}

\maketitle

\section{\label{sec:intro}Introduction}
The generation of runaway electrons has in recent years emerged as one of the major challenges to address before fusion can become a scientifically and economically viable energy source. 
Electron runaway is the phenomenon of electrons being continuously accelerated to relativistic speeds by an electric field~\cite{wilson1925,dreicer1959, dreicer1960}. 
In a tokamak, runaway electrons can be generated during sudden plasma-terminating events called disruptions~\cite{helander2002}, mainly caused by current-driven instabilities~\cite{boozer2012} or the influx of radiating impurities~\cite{breizman2019}. 
During a tokamak disruption, the generated current of runaway electrons can be sufficiently large to risk seriously damaging the plasma-facing wall components upon loss of control of the plasma position~\cite{ratynskaia2025,rizzi2026}. 
The rate of runaway production is exponentially sensitive to the magnitude of the initial plasma current~\cite{rosenbluth1997}, making generation of runaway electrons a growing concern for reactor-scale tokamaks~\cite{ekmark2024, fil2024, ekmark2025, sweeney2026}. 

Historically, runaway electrons have not been considered a concern in stellarators, but this could change as we approach reactor-relevant scales. 
Fundamentally, a stellarator is not reliant on an externally driven plasma current for producing the rotational transform required for plasma confinement~\cite{helander2014, boozer2021}. 
Due to the absence of an externally driven plasma current and consequently the absence of disruptions caused by current-driven instabilities~\cite{helander2012}, runaway generation has not been anticipated to be an issue in stellarators and runaway electrons in stellarators constitutes a largely unexplored area of research. 
However, plasma currents of considerable magnitude driven by gradients in plasma density and temperature (so called bootstrap currents) can be present in stellarators~\cite{helander2017, landreman2022}. 
Bootstrap currents of several mega-amperes are predicted for quasi-axisymmetric stellarators. 
Quasi-axisymmetry is a subset of quasi-symmetry~\cite{landremanPaul2022}, i.e.~the property of the magnetic field strength being symmetric in a special coordinate system~\cite{boozer1999}, 
which achieves good particle confinement, a property that is essential for reactors~\cite{landreman2012}. 
For instance, the bootstrap current in the conceptual quasiaxisymmetric stellarator power plant ARIES-CS was predicted to be \SI{4}{MA}~\cite{najmabadi2008}. 
As stellarator development is also approaching reactor-relevant scales~\cite{gates2017, hegna2022, warmer2024}, questions about runaway electrons in stellarators are emerging. 
Can there be off-normal events in stellarators that facilitate runaway generation? 
How severe can the runaway generation in a stellarator be? 
Will runaway electrons pose the same challenges in stellarators as they do in tokamaks? 

Even though disruptions induced by current-driven instabilities are not a concern for stellarators, there are other conceivable events that could trigger facilitation of runaway electron generation. 
A recent paper investigated runaway electron generation enabled by coil quenching, whereby a toroidal electric field can be induced, and found that electron runaway is theoretically possible during such events~\cite{aleynikov2026}. 
From another perspective, if a temperature collapse (and subsequent current decay) could be triggered in a stellarator, this could enable runaway generation under similar circumstances as in a tokamak disruption. 
During a temperature collapse, an electric field would be induced to maintain the total plasma current, which could be sufficiently large for runaway generation. 
For instance, radiation-driven temperature collapse triggered by impurities entering the plasma~\cite{dinklage2019} is one imaginable precursor of considerable runaway generation in a stellarator. 

If there is a plasma current in a stellarator, it will pose the same potential problem for avalanche runaway generation as in a tokamak. 
In Ref.~\cite{rosenbluth1997}, it was demonstrated that the magnitude of the plasma current affects the avalanche generation of runaway electrons in a tokamak. 
In fact, since the avalanche multiplication factor is exponentially sensitive to the plasma current, it is the most concerning runaway electron generation mechanism in reactor-scale tokamaks. 
The reasoning of Ref.~\cite{rosenbluth1997} is importantly also valid in a stellarator with a decaying bootstrap current, indicating that runaway electron generation through avalanching could also be of concern in stellarators.

In this paper, we present one of the first studies of runaway electrons in stellarators, by investigating runaway generation during temperature collapse scenarios in a stellarator configuration with varying plasma current. 
To cover a varied set of possible stellarator temperature collapse scenarios, we explore how the temperature collapse time scale, post-collapse temperature, and magnitude of plasma current influence the generated runaway current. 
The exploration is conducted through simulations using the runaway electron simulation tool \textsc{Dream}~\cite{dream}, which has been upgraded to accurately model stellarator plasmas.  
We demonstrate that significant runaway electron generation is possible in stellarators and identify under which conditions it could be a concern.

\section{\label{sec:physmod}Numerical model}
To study runaway electrons in stellarators, the numerical model in the simulation tool \textsc{Dream} has been extended to be applicable for stellarator configurations as well. 
\textsc{Dream} is a simulation tool developed for studying runaway electrons in tokamaks at a relatively low computational cost. 
To accurately describe the runaway electron evolution, the plasma is simulated self-consistently through coupled differential equations in one spatial dimension. 
\textsc{Dream} reduces the dimensionality to one spatial dimension by using flux surface averaged quantities and resolving the plasma spatially through the flux surface label coordinate. 
The flux surface averaging in \textsc{Dream} has been generalized to be applicable to stellarators, as they are not toroidally symmetric like tokamaks. 
A flux surface average of quantity $X$ is defined by
\begin{align}
    \left\langle X\right\rangle&=\frac{1}{V'}\int_0^{2\pi}\dd\varphi \int_{-\pi}^\pi\dd\theta \sqrt{g} X,\\
    V'&=\int_0^{2\pi}\dd\varphi \int_{-\pi}^\pi\dd\theta \sqrt{g} =\pdv{V}{r},
\end{align}
where $\varphi$ is the toroidal and $\theta$ the poloidal angle coordinate, $V$ is the volume enclosed by the radial coordinate $r$ and $\sqrt{g} =1/\abs{\nabla\varphi\cdot(\nabla\theta\times\nabla r)}$ is the spatial Jacobian. 
The radial coordinate $r$ is defined by the flux surface label $\rho$ and average minor radius $a$ as $r=a\rho$. 
\textsc{Dream} allows for simulating in up to two dimensions in momentum space in tokamaks, but the currently implemented model for stellarators only allows fluid simulations. 
Furthermore, considerations have also had to be made for the implementations of Faraday's and Ampère's laws. 

A form of Faraday's law applicable to stellarators and tokamaks is given in Ref.~\cite{strand2001} as
\begin{equation}
    \left\langle \boldsymbol{E}\cdot\boldsymbol{B} \right\rangle V' 
    = \pdv{\psi_{\rm p}}{t}\pdv{\psi_{\rm t}}{r}
    -\pdv{\psi_{\rm t}}{t}\pdv{\psi_{\rm p}}{r},
    \label{eq:farorg}
\end{equation}
where $\boldsymbol{E}$ is the electric field, $\boldsymbol{B}$ the magnetic field, $\psi_{\rm t}$ the toroidal magnetic flux and $\psi_{\rm p}$ the poloidal magnetic flux driven by the loop voltage through 
\begin{align}
    \left.\pdv{\psi_{\rm p}}{t}\right|_{r} \equiv V_{\rm loop}(r).
\end{align} 
The toroidal flux can be assumed to be approximately constant in time if the magnetic coils are unaffected during the temperature and current decays, which simplifies the equation to 
\begin{equation}
    \left\langle \boldsymbol{E}\cdot\boldsymbol{B} \right\rangle V' 
    = \pdv{\psi_{\rm p}}{t}\pdv{\psi_{\rm t}}{r}.
    \label{eq:farsimp}
\end{equation}
The toroidal flux partial derivative $\partial \psi_{\rm t}/\partial r$ can be obtained from the rotational transform
\begin{align}
    \iota &= 
    \frac{\left\langle \boldsymbol{B}\cdot\nabla\theta \right\rangle}
    {\left\langle \boldsymbol{B}\cdot\nabla\varphi \right\rangle}
    = \pdv{\psi_{\rm p}}{\psi_{\rm t}}
    =\frac{\partial\psi_{\rm p} / \partial r}{\partial\psi_{\rm t} / \partial r}
    \label{eq:iota}
\end{align}
where we can use
\begin{align}
    \left\langle \boldsymbol{B}\cdot\nabla\theta\right\rangle
    =\frac1{V'}\int_{0}^{2\pi}\dd\varphi\int_{-\pi}^{\pi}\dd\theta
    \sqrt{g} \boldsymbol{B}\cdot\nabla\theta
    =\frac{2\pi}{V'}\pdv{\psi_{\rm p}}{r},
    \label{eq:btheta}
\end{align}
since $\psi_{\rm p}=\int_{0}^{2\pi}\dd\varphi\int_{0}^{r}\dd r' \sqrt{g} \boldsymbol{B}\cdot\nabla\theta+\int V_{\rm loop}(0) {\rm d} t$. 
Combining equations \eqref{eq:iota} and \eqref{eq:btheta}, we obtain
\begin{align}
    \pdv{\psi_{\rm t}}{r}
    =\frac{V'}{2\pi}\left\langle \boldsymbol{B}\cdot\nabla\varphi \right\rangle,
\end{align}
which together with equation \eqref{eq:farsimp} yields
\begin{align}
    \pdv{\psi_{\rm p}}{t}
    &=\frac{2\pi\left\langle \boldsymbol{E}\cdot\boldsymbol{B} \right\rangle}
    {\left\langle \boldsymbol{B}\cdot\nabla\varphi \right\rangle}. 
\end{align}
Notably, it is the same form of Faraday's law as is already used in \textsc{Dream} for tokamaks. 

A similarly general form for Ampère's law can also be found in Ref.~\cite{strand2001} as
\begin{align}
    \begin{split}
        \mu_0 I(r,t) &=
        \frac{V'}{4\pi^2}
        \left\langle \frac{g_{\theta\theta}}{g} \right\rangle
        \pdv{\psi_{\rm p}}{r}\\
        &\hspace{0.8cm}+\frac{V'}{4\pi^2}
        \left\langle \frac{g_{\theta\varphi}(1+\lambda_\theta) - g_{\theta\theta}\lambda_\varphi}{g} \right\rangle
        \pdv{\psi_{\rm t}}{r},
        \label{eq:AmpSt}
    \end{split}
\end{align}
where $g_{ij}$ are elements of the covariant metric tensor and $\lambda_i=\partial \lambda/\partial i$ are derivatives of the poloidal magnetic stream function $\lambda$ for obtaining the straight-field-line poloidal angle $\vartheta = \theta + \lambda$ associated with $\varphi$.
Furthermore, $I(r,t)$ is the current enclosed by the flux surface $r$ at time $t$, i.e. 
\begin{align}
    I(r,t)&=\frac1{2\pi}\int_{0}^{2\pi}\left[\int \boldsymbol{J}\cdot\dd\boldsymbol{S}\right]\dd\varphi. 
\end{align}
We can substitute the surface element $\dd\boldsymbol{S}=\pdv{\boldsymbol{x}}{r}\times\pdv{\boldsymbol{x}}{\theta}\dd \theta\dd r$ by using the definition $\nabla\varphi=\frac{1}{\sqrt{g}}\left[\pdv{\boldsymbol{x}}{r}\times\pdv{\boldsymbol{x}}{\theta}\right]$ and obtain
\begin{align}
    I(r,t)&=\frac1{2\pi}\int_{0}^{2\pi}\dd\varphi\int_0^r\dd r\int_{-\pi}^{\pi}\ \dd\theta\boldsymbol{J}\cdot\nabla\varphi\sqrt{g} ,
\end{align}
which contains the two integrals used for flux surface averaging. 
Using the definition of the current density $\boldsymbol{J}=j_\parallel\boldsymbol{B}/B$, and assuming that $j_\parallel/B$ is approximately constant on a flux surface, we arrive at the expression
\begin{align}
    I(r,t)&=\frac1{2\pi}\int_0^r V'\left\langle \boldsymbol{B}\cdot\nabla\varphi\right\rangle\frac{j_\parallel}{B}\dd r.
\end{align}
That the quantity $j_\parallel/B$ is an invariant quantity on a flux surface is a good approximation for quasi-symmetric stellarators with small relative variation in the magnetic field strength. 
Finally, by differentiating equation \eqref{eq:AmpSt} with respect to $r$, we obtain 
\begin{align}
    \begin{split}
        &2\pi\mu_0 \left\langle \boldsymbol{B}\cdot\nabla\varphi\right\rangle\frac{j_\parallel}{B}
        =\\
        &=\frac{1}{V'}
        \pdv{r}\left[V'\left\langle \frac{g_{\theta\theta}}{g} \right\rangle
        \pdv{\psi_{\rm p}}{r}\right.\\
        &\hspace{1.8cm}\left.+
        V'\left\langle \frac{g_{\theta\varphi}(1+\lambda_\theta) - g_{\theta\theta}\lambda_\varphi}{g } \right\rangle
        \pdv{\psi_{\rm t}}{r}
        \right].\label{eq:ampere}
    \end{split}
\end{align}
For a tokamak, $\frac{g_{\theta\theta}}{g}=\frac{\abs{\nabla r}^2}{R^2}$ and ${g_{\theta\varphi}(1+\lambda_\theta) - g_{\theta\theta}\lambda_\varphi=0}$, reducing equation \eqref{eq:ampere} to the form implemented for tokamaks in \textsc{Dream}. 
To obtain the magnetic configuration data for a stellarator necessary for evolving the simulated plasma -- e.g~the metric tensor components, magnetic field strength and toroidal magnetic flux -- we utilize the DESC~equilibrium solver~\cite{dudt2020}. 
Since our numerical model uses general expressions for Faraday's and Ampère's laws valid for both tokamaks and stellarators, we have validated it against the original numerical model in \textsc{Dream} for a tokamak disruption and the two models demonstrate good agreement.

\section{\label{sec:simset}Simulation set-up}
Runaway electron generation is sensitively dependent on the electron temperature evolution and magnitude of the initial plasma current. 
The electron temperature affects both the seed generation through the rate of the temperature decay and the avalanche generation through the post-collapse temperature.
Avalanche generation also sensitively depends on the size of the initial plasma current~\cite{rosenbluth1997,hesslow2019}, as discussed in section \ref{sec:intro}. 
As a result, the risk of significant runaway electron generation in stellarators is case-dependent, both with regards to plasma and impurity composition as well as stellarator configuration. 

\begin{figure}
    \centering
    \includegraphics[width=8.64cm]{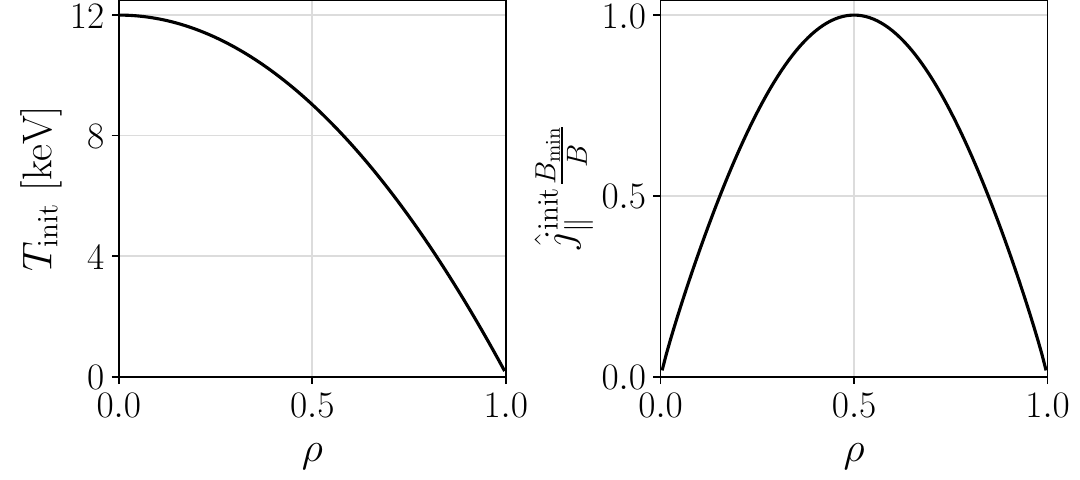}
    \put(-143,99){(a)}
    \put(-20,99){(b)}
    \caption{Initial profiles for (a) the plasma temperature, and (b) the current density, plotted against the normalized minor radius $\rho$. }
    \label{fig:initprof}
\end{figure}

To thoroughly study runaway generation during radiative collapse in stellarators, we scan over temperature decay time scale, post-collapse temperature, and initial plasma current. 
The plasma currents considered are $I_{\rm p}\in[0.1,\ 10]\,\si{MA}$, where \SI{0}{MA} would correspond to a stellarator configuration where all of the rotational transform is generated by coils and \SI{10}{MA} to a more tokamak-like configuration. 
As for the temperature evolution, it is prescribed as an exponential decay, where we parametrize the decay time scale by $\tau\in[0.02,\ 2]\,\si{ms}$, where the shorter time scales correspond to stronger hot-tail generation, and the post-collapse temperature by $T_{\rm fin}\in[1,\ 100]\,\si{eV}$, where the lower temperatures correspond to stronger avalanche generation. 
The temperature evolution is thus evolved according to 
\begin{equation}
    T(r, t)=(T_{\rm init}(r)-T_{\rm fin})\exp(-t/\tau)+T_{\rm fin},
    \label{eq:expdec}
\end{equation}
where the initial temperature radial profile $T_{\rm init}(r)$ is illustrated in figure \ref{fig:initprof}a. 
The post-collapse temperature can be determined from plasma composition, but is treated as a free parameter since we have excluded impurities in the simulations. 
By performing these scans over temperature evolution and plasma current, we thus intend to present a comprehensive study of the risk of runaway electron generation in stellarators, relevant for both a wide range of  configurations and temperature collapse scenarios. 

While the temperature evolution is modelled as exponential decay according to equation \eqref{eq:expdec}, the ion densities, current density, poloidal flux and electric field are evolved self-consistently. 
The total current density, poloidal flux and electric field are evolved according to Faraday's and Ampère's laws, as described in section \ref{sec:physmod}, while the Ohmic current density is determined through Ohm's law. 
As described above, we have varied the plasma current, but the initial current density profile has been kept fixed and is illustrated in figure \ref{fig:initprof}b. 
The current density profile was chosen to vanish at the plasma centre and edge to emulate typical features of a bootstrap current, since the plasma current is mainly driven by pressure gradients during normal operation in a stellarator. 
Important to note however is that, while the current density evolution is self-consistent with regards to the electric field, poloidal flux and temperature, the initial current density profile is not consistent with the plasma pressure in the simulations, as described by Ref.~\cite{landreman2022} for quasi-axisymmetric profiles. 
Consistency between the plasma pressure and current density profile is not enforced for simplicity, and to more easily isolate the effect of varying the plasma current and current density. 
The radial profiles of the deuterium and tritium densities have been set to be constant, for simplicity, at \SI{1e20}{m^{-3}} each, and the electron density is evolved to ensure net neutrality. 
Furthermore, the ions are evolved according to ionization and recombination rates from the AMJUEL-database~\cite{amjuel}. 
The code used and the data produced can be found at Zenodo~\footnote{I.~Ekmark and M.~Landreman, Runaway electrons in stellarators: Unlikely or unavoidable? [Dataset], Zenodo, 2026, \href{https://doi.org/10.5281/zenodo.23672439}{https://doi.org/10.5281/zenodo.23672439}.}. 

\begin{figure}
    \centering
    \includegraphics[width=8.64cm]{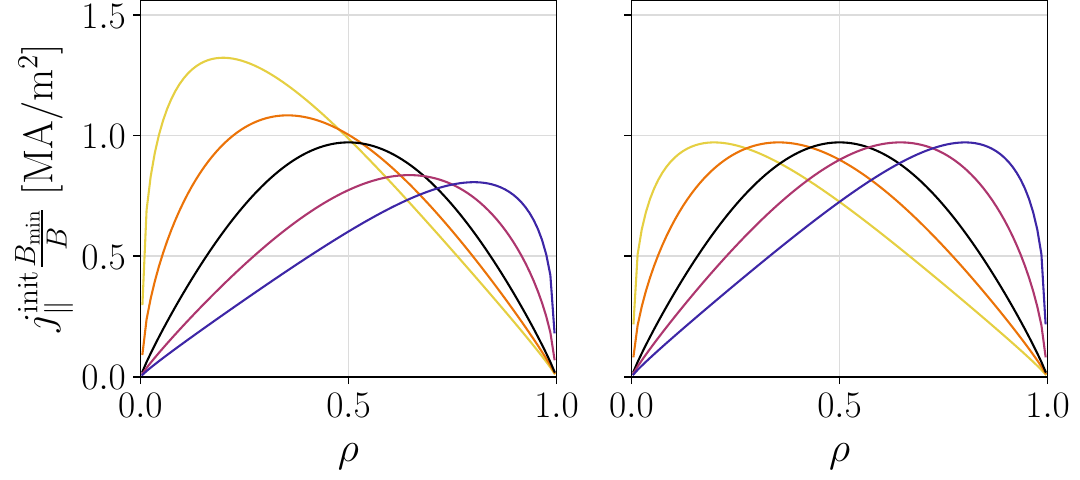}
    \put(-213,99){(a)}
    \put(-101,99){(b)}
    \caption{Initial current densities used to study profile sensitivity (a) with a constant plasma current, and (b) with a constant maximum current density. }
    \label{fig:initprofBS}
\end{figure}

The peak of the current density profile has been set at $\rho=0.5$, i.e.~at half the plasma minor radius, for most of the simulations, but sensitivity scans of the peak location have also been performed. 
However, both the magnitude of the current density and the total plasma current have a strong effect on the runaway electron generation, and it is impossible to keep them both fixed while varying the current density profile shape. 
Therefore, we have done sensitivity scans for a fixed initial plasma current of \SI{6}{MA}, with profiles as illustrated in figure \ref{fig:initprofBS}a, and for a fixed maximum current density of \SI{1}{MA/m^2}, as illustrated in figure \ref{fig:initprofBS}b. 
For the sensitivity scan with fixed maximum current density, the sample with peak at $\rho=0.5$ has a plasma current of \SI{6}{MA}, and is identical to the sample at $\rho=0.5$ for the fixed initial plasma current case.

\begin{figure}
    \centering
    \includegraphics[width=4.8cm]{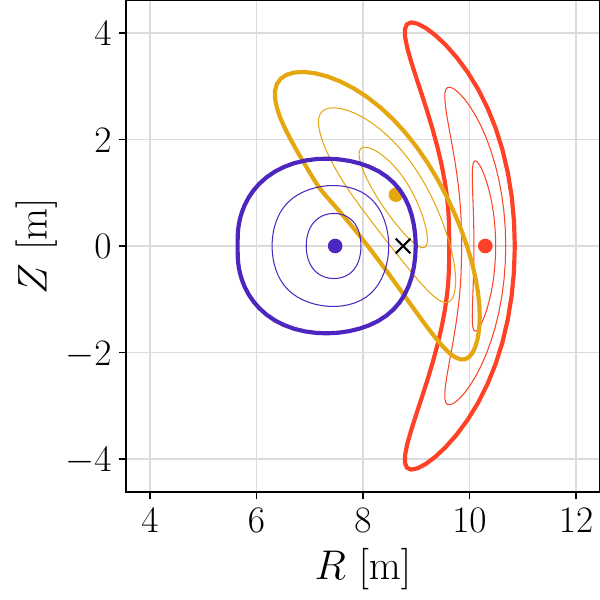}
    \caption{Stellarator configuration used for the simulations with two field periods. The black cross marks the major radius in the $Z=0$-plane, while the flux surfaces for $\varphi\in\{0,\ \pi/4,\ \pi/2\}$ are plotted in red, yellow and purple, respectively.}
    \label{fig:config}
\end{figure}

We have simulated a deuterium-tritium plasma in a fixed quasi-axisymmetric stellarator configuration illustrated in figure \ref{fig:config}. 
Note that the magnetic configuration is not consistent with the plasma current densities used, but has instead been kept fixed to isolate effects of changing the other parameters. 
The configuration is similar to the quasi-axisymmetric configuration in appendix C of Ref.~\cite{landreman2022}, but has an aspect ratio $A=5$ instead of $A=6$~\footnote{S.~Buller, Hifistell finite-beta QA proposal 1, Sec.~1.2 ($A = 5$), 2024, \href{https://sb0095.mycpanel.princeton.edu/QA/notes.html}{https://sb0095.mycpanel.princeton.edu/QA/notes.html}.}. 
It has a major radius $R_0=\SI{8.8}{m}$ and minor radius of $a=\SI{1.7}{m}$, with two field periods and an average rotational transform $\iota=0.42$. 
We have assumed that the plasma is surrounded by a perfectly conducting, closed toroidal structure at minor radius $b$. 
The assumption of a perfectly conducting wall does affect the plasma evolution and runaway generation, as having a finite resistive wall time enables the electric field diffusion across the surrounding structures and in turn increases the avalanche. 
Including a more detailed analysis of the effect of a finite wall time would introduce other complications requiring more careful consideration outside of the scope of this paper. 
We therefore leave such analyses to future studies, and keep the assumption of an infinite wall time, as has been done in previous, similar studies e.g.~in Helios~\cite{swanson2026}. 
Importantly, the distance $b$ from the plasma to the toroidally closed conducting structure determines the avalanche gain through setting the poloidal magnetic flux, and is thus a sensitive free parameter. 
To get a unique solution when evolving the magnetic flux through Ampère's law (equation \eqref{eq:ampere}), the magnetic flux at the wall is used as a boundary condition. 
The poloidal flux at the plasma edge
$\psi_{\rm p}(a,t)=\psi_{\rm p}(b,t) - I(a,t)R_0\mu_0\ln \frac{b}{a}$, 
where $R_0\mu_0\ln \frac{b}{a}$ represents an approximate edge-wall mutual flux inductance which is valid in the cylindrical limit. 
Note that for a perfectly conducting wall $\psi_{\rm p}(b)=0$. 
For most of the simulations, we have assumed the distance between the magnetic axis and the toroidally conducting structure $b=1.1a$. 
To study the sensitivity however, we have also run simulations for $b/a\in[1,\ 1.5]$.

During the temperature collapse of a deuterium-tritium plasma, all runaway electron generation mechanisms relevant for a fusion plasma can be active. 
The generation mechanisms relevant for a fusion plasma are Dreicer~\cite{hesslow2019dreicer}, hot-tail~\cite{smith2008} and avalanche generation~\cite{hesslow2019}, as well as generation from tritium beta decay and Compton scattering induced by photons emitted from the activated walls. 
The models for describing runaway generation from tritium beta decay and Compton scattering are detailed in Ref.~\cite{vallhagen2020}. 
For the generation from Compton scattering, a gamma photon flux spectrum is needed, and we have used a reactor-relevant spectrum, specifically the spectrum for the ITER-tokamak presented in Ref.~\cite{martinsolis2017} due to it being reactor-relevant. 
Aside from the generation mechanisms, the runaway electron population can also be influenced by radial transport. 
Since transport of runaway electrons generally leads to a reduction of the runaway electron population, we have chosen to neglect this effect to give conservative, that is maximum, predictions of the runaway electron densities.

\section{\label{sec:res}Results}
In this section, we present the results of the simulations of stellarator temperature and current collapse using the extended \textsc{Dream} simulation tool. 
As described in section \ref{sec:simset}, we have studied how the runaway electron generation is affected by the initial plasma current $I_{\rm p}$, post-collapse temperature $T_{\rm fin}$ and temperature decay time scale $\tau$. 
However, when referencing the temperature collapse time scale, we will instead use the thermal quench time $t_{\rm TQ}$ for simplicity, and we define $t_{\rm TQ}$ as the time when $T(t_{\rm TQ})=T_{\rm fin}+\SI{1}{eV}$. 
We have also studied how the runaway generation is affected by the bootstrap current profile and distance to the toroidally closed conducting structure.

\begin{figure}
    \centering
    \includegraphics[width=8.64cm]{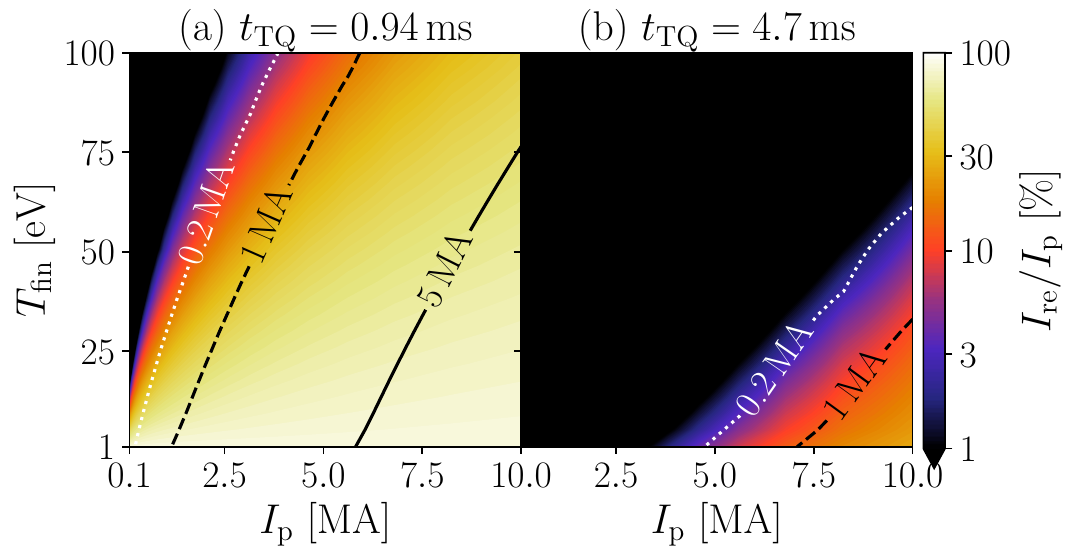}
    \caption{Fraction of initial plasma current that is converted to runaway current as a function of initial plasma current and post-collapse temperature for a thermal quench time of (a) $t_{\rm TQ}=\SI{0.94}{ms}$ (${{\tau=\SI{0.1}{ms}}}$) and (b) $t_{\rm TQ}=\SI{4.7}{ms}$ (${\tau=\SI{0.5}{ms}}$). 
    Contours of ${I_{\rm re}\in\{0.2,\  1,\ 5\}\,\si{MA}}$ are indicated by dotted, dashed and solid lines, respectively.}
    \label{fig:REfrac}
\end{figure}

As presented in figure \ref{fig:REfrac}, the severity of the runaway generation varies significantly depending on initial plasma current $I_{\rm p}$, post-collapse temperature $T_{\rm fin}$ and thermal quench time $t_{\rm TQ}$. 
The figure depicts the percentage of the initial plasma current that is converted into runaway current as a function of initial plasma current $I_{\rm p}$ and post-collapse temperature $T_{\rm fin}$ for thermal quench time $t_{\rm TQ}\in\{0.94,\ 4.7\}\,\si{ms}$.
For fast temperature decays ($t_{\rm TQ}=\SI{0.94}{ms}$), all but the upper left corner, corresponding to low $I_{\rm p}$ and high $T_{\rm fin}$, displays more than \SI{10}{\percent} conversion, resulting in mega-ampere runaway currents for most $I_{\rm p}>\SI{1}{MA}$. 
On the contrary, for slower temperature decays ($t_{\rm TQ}=\SI{4.7}{ms}$), a negligible fraction of the initial plasma current is converted into runaway current for all but the highest $I_{\rm p}$ and lowest $T_{\rm fin}$, and even then, the conversion is less than \SI{25}{\percent}. 
It is expected that the runaway generation is enhanced with a shorter decay time scale since it determines the severity of the hot-tail generation. 
At $t_{\rm TQ}=\SI{0.94}{ms}$, the hot-tail generation is strong and the conversion fraction less sensitive to $I_{\rm p}$, while the dependence on $I_{\rm p}$ is stronger at $t_{\rm TQ}=\SI{4.7}{ms}$, where the avalanche generation is more important. 

\begin{figure}
    \centering
    \includegraphics[width=8.64cm]{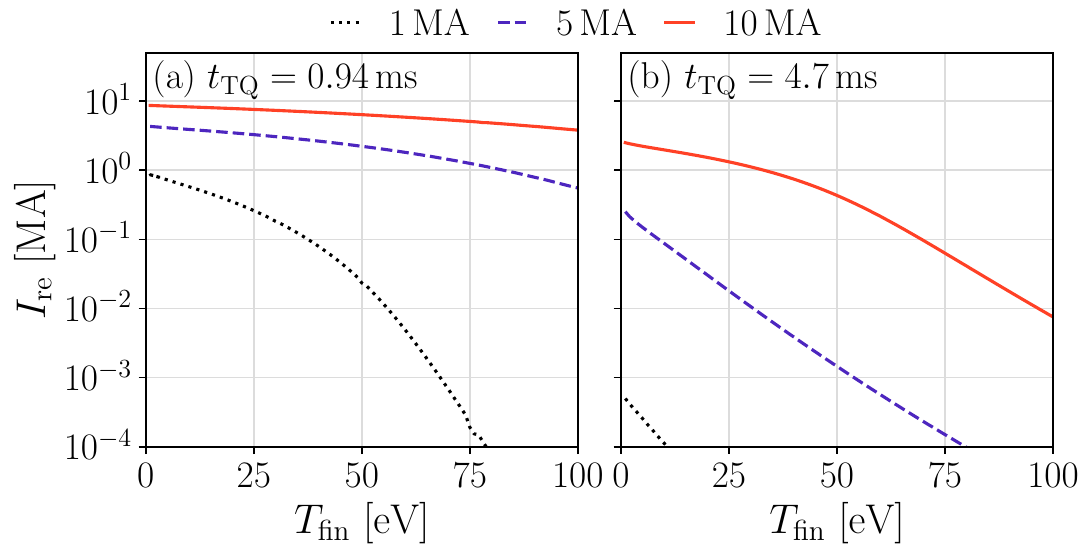}
    \put(-185,119){$I_{\rm p}$:} 
    \caption{Generated runaway current as a function of post-collapse temperature $T_{\rm fin}$ for a thermal quench time of (a) $t_{\rm TQ}=\SI{0.94}{ms}$ (${{\tau=\SI{0.1}{ms}}}$) and (b) $t_{\rm TQ}=\SI{4.7}{ms}$ (${\tau=\SI{0.5}{ms}}$). The different curves, of varying colours, correspond to different initial plasma currents $I_{\rm p}$.  }
    \label{fig:TendTrend}
\end{figure}
More important than the current conversion rate is the magnitude of the generated runaway current -- in a tokamak it should be $\lesssim\SI{100}{kA}$ ideally~\cite{lehnen2021} -- which can be quite sensitive to both the initial plasma current and post-collapse temperature, as illustrated in figure \ref{fig:TendTrend}. 
In the figure, the runaway current $I_{\rm re}$ is plotted as a function of the post-collapse temperature $T_{\rm fin}$ for initial plasma current $I_{\rm p}\in\{1,\ 5,\ 10\}\,\si{MA}$ (curves of varying colours) and thermal quench time $t_{\rm TQ}\in\{0.94,\ 4.7\}\,\si{ms}$ (corresponding to \ref{fig:TendTrend}a and \ref{fig:TendTrend}b, respectively).  
The runaway current is invariably sensitive to the initial plasma current since it determines the avalanche gain, and sets an upper limit for how large runaway current can be generated. 
Also the post-collapse temperature $T_{\rm fin}$ influences the avalanche multiplication, through the conductivity and subsequently the electric field. 
It has a strong effect on the generated runaway current if the temperature decay is slow ($t_{\rm TQ}\geq\SI{4.7}{ms}$) or if the plasma current is low ($I_{\rm p}\lesssim\SI{1}{MA}$). 
More specifically, figure \ref{fig:TendTrend} suggests that significant runaway currents can be generated in stellarators if two out of three of the following conditions are met: (1) large $I_{\rm p}$ ($\gtrsim\SI{5}{MA}$), (2) low $T_{\rm fin}$ ($\lesssim\SI{20}{eV}$), and (3) short $t_{\rm TQ}$ ($\lesssim\SI{2.8}{ms}$). 
As most stellarators have small plasma currents, this implies that for significant runaway currents to be generated in such devices, severe thermal quenches -- that is fast and with low post-collapse temperatures -- are required. 
However; less extreme thermal quenches might still generate a considerable runaway current in stellarators with plasma currents of several mega-amperes, such as quasi-axisymmetric stellarators.

\begin{figure}
    \centering
    \includegraphics[width=8.64cm]{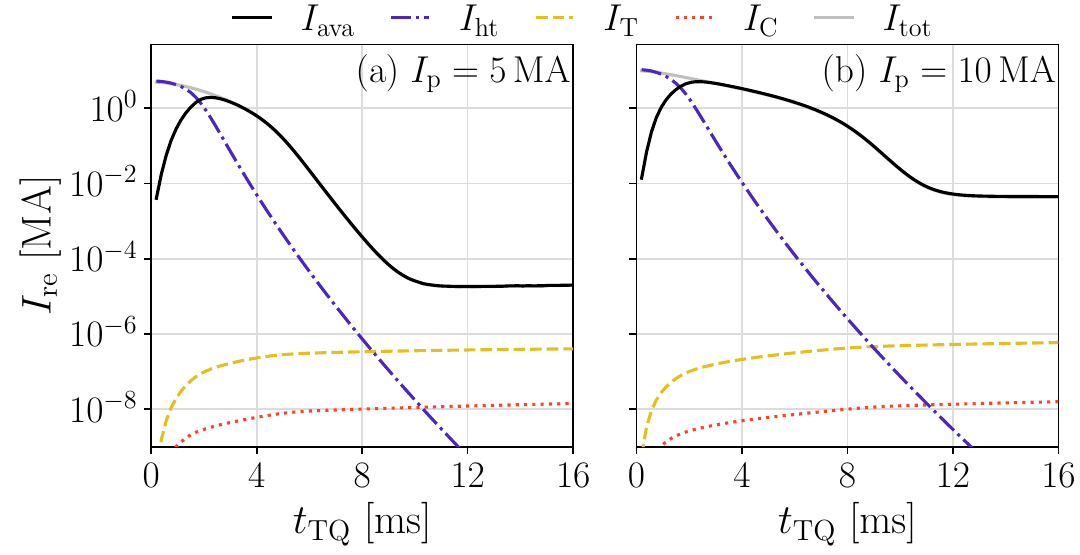}
    \caption{The runaway current generated by avalanche $I_{\rm ava}$ (black solid), hot-tail $I_{\rm ht}$ (purple dash-dotted), tritium beta decay $I_{\rm T}$ (yellow dashed), and Compton scattering $I_{\rm C}$ (red dotted) as a function of the thermal quench time $t_{\rm TQ}$ for (a) $I_{\rm p}=\SI{5}{MA}$ and (b) $I_{\rm p}=\SI{10}{MA}$. Dreicer generation is negligible. All simulations correspond to $T_{\rm fin}=\SI{1}{eV}$.}
    \label{fig:mechs}
\end{figure}
Figure \ref{fig:mechs} reveals that the main generation mechanisms driving the runaway generation during such a stellarator temperature collapse are hot-tail and avalanche generation, and that the dynamics are sensitively dependent on the thermal quench time. 
The figure presents the runaway currents $I_{\rm re}$ generated by the different mechanisms: avalanche, hot-tail, tritium beta decay and Compton scattering. 
Dreicer generation is not included since it was negligible compared to the other generation mechanisms in these simulations. 
Since the avalanche generation is affected by how low the post-collapse temperature is, we have used $T_{\rm fin}=\SI{1}{eV}$ to be conservative. 
As demonstrated in the figure, the dynamics can be divided into three regimes based on the thermal quench time. 
For very fast temperature collapses ($t_{\rm TQ}\lesssim\SI{2}{ms}$), hot-tail generation is the dominating generation mechanism and there is almost full conversion of the initial plasma current to runaway current. 
Hot-tail generation is still the dominating seed mechanism for intermediate temperature collapses ($2\lesssim t_{\rm TQ}\lesssim\SI{9}{ms}$), but avalanche generation is the dominating mechanism overall. 
For slower temperature collapses ($t_{\rm TQ}\gtrsim\SI{9}{ms}$), tritium beta decay dominates the seed generation and the generation stabilizes. 
The generation from tritium beta decay is facilitated by the low post-collapse temperature, which induces a strong electric field. 
With a higher post-collapse temperature, generation from Compton scattering is the dominating seed mechanism for slower temperature collapses. 
As $\tau$ is increased beyond \SI{2}{ms} -- corresponding to $t_{\rm TQ}>\SI{20}{ms}$ -- the avalanche generation decreases slowly. 

Figure \ref{fig:mechs} also illustrates that the total runaway current generated varies over orders of magnitude for $t_{\rm TQ}\in[0.2,10]\,\si{ms}$, for both \SI{5}{MA} (\ref{fig:mechs}a) and \SI{10}{MA} (\ref{fig:mechs}b) of initial plasma current.
If $t_{\rm TQ}<\SI{9}{ms}$, a significant runaway current can be generated ($I_{\rm re}\gtrsim\SI{100}{kA}$), facilitated by the strong hot-tail generation. 
When hot-tail generation is comparable to or less than the generation from tritium beta decay, the runaway current is $\lesssim\SI{10}{kA}$. 
In a similar simulation set-up for an ITER-like configuration with $I_{\rm p}=\SI{10}{MA}$, the avalanche is sufficiently strong to generate a $\sim\SI{1}{MA}$ runaway current even from a seed of runaways generated only by tritium beta decay. 
On the contrary, in our stellarator simulations, both the tritium-beta-decay seed and avalanche generation rates are lower than in the ITER-like configuration, implying that the runaway generation is significantly weaker in a stellarator even with a plasma current of the same magnitude.

\begin{figure}
    \centering
    \includegraphics[width=8.64cm]{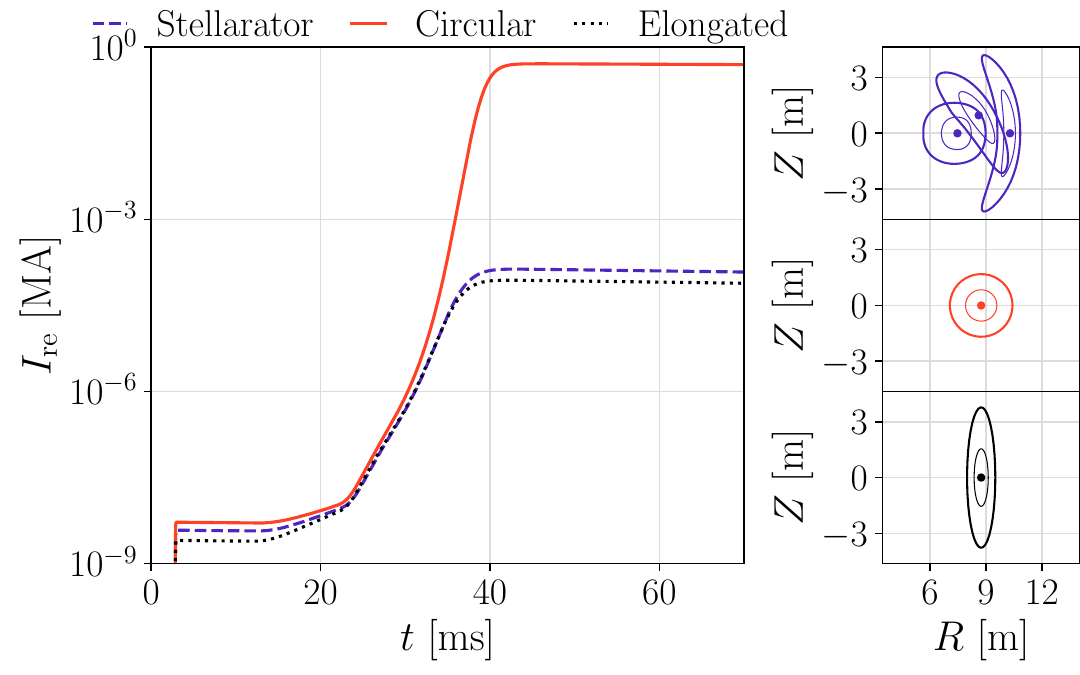}
    \put(-210,130){(a)}
    \put(-45,135){(b)}
    \put(-45,95.25){(c)}
    \put(-45,55.5){(d)}
    \caption{(a) Runaway current evolution for a stellarator plasma (dashed purple), and circular (solid red) and elongated (dotted black) tokamak plasmas. The cross sections for the (b) stellarator, (c) circular, and (d) elongated plasmas used are also illustrated. }
    \label{fig:Ireshapes}
\end{figure}
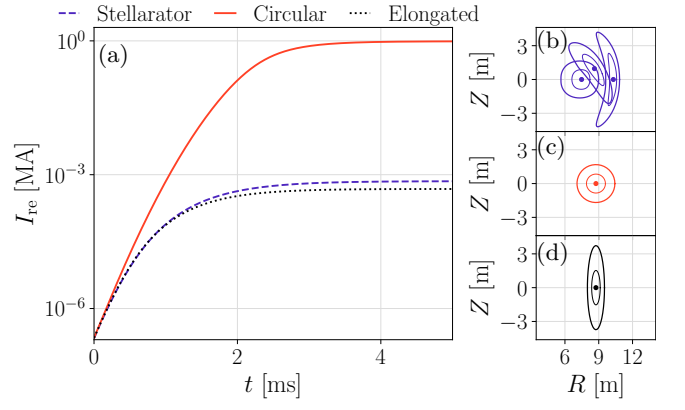

Our simulations thus predict the seed generation and avalanche multiplication to be weaker in a stellarator compared with a tokamak, even for stellarators of high plasma currents, which can be explained by the more elongated shape of stellarator configurations. 
Fülöp \emph{et al.}~showed that in tokamaks, elongation of the plasma reduces the electric field diffusion and in turn both the seed and avalanche generation rates. 
More specifically, the maximum avalanche gain is reduced by a factor $2/(\kappa+\kappa^{-1})$, with $\kappa$ parametrizing the elongation~\cite{fulop2020}. 
The elongation of stellarator plasmas is not as easily parametrizable as that of tokamak plasmas due to the toroidal variation of the cross-section, but the underlying effect on the avalanche should be similar. 
This effect can be demonstrated by comparing the same, slow-TQ scenario ($\tau=\SI{4}{ms}$ and $T_{\rm fin}=\SI{1}{eV}$) in a stellarator plasma (figure \ref{fig:Ireshapes}b) to a circular (figure \ref{fig:Ireshapes}c) and elongated (figure \ref{fig:Ireshapes}d) tokamak plasma, all with initial plasma current $I_{\rm p}=\SI{10}{MA}$ and plasma volume $V=\SI{490}{m^3}$. 
The elongation $\kappa(\rho)$ of the tokamak plasma was chosen so that 
\begin{equation}
    \frac{1}{a}\left[V'\left\langle \frac{\abs{\nabla r}^2}{R^2}\right\rangle\right]_\kappa \Bigg|_{\rho}^{\rm elongated}=
    \frac{V'}{a}\left\langle \frac{g_{\theta\theta}}{g} \right\rangle\Bigg|_{\rho}^{\rm stellarator} 
\end{equation}
for normalized minor radius $\rho\in[0,1]$, while also altering the plasma minor radius of the elongated plasma to keep the volume fixed. 
Figure \ref{fig:Ireshapes}a shows that the avalanche generation is similar for the stellarator and elongated plasmas, with the generated runaway current being $\SI{140}{A}$ and $\SI{90}{A}$, respectively. 
The difference in runaway current can be explained by the functions describing the current $I(\rho,t=0)$ enclosed within the flux surface at $\rho$ (see equation \eqref{eq:AmpSt}) being slightly different. 
For the circular plasma, on the other hand, a \SI{520}{kA} runaway current was generated, i.e.~more than $3\,000$ times more runaways were generated than in the strongly shaped plasmas, mainly due to the increased avalanche generation rate. 
This demonstrates that the elongation of a stellarator plasma will reduce the seed generation and avalanche multiplication and can have a strong effect on the runaway dynamics. 

\begin{figure}
    \centering
    \includegraphics[width=7.2cm]{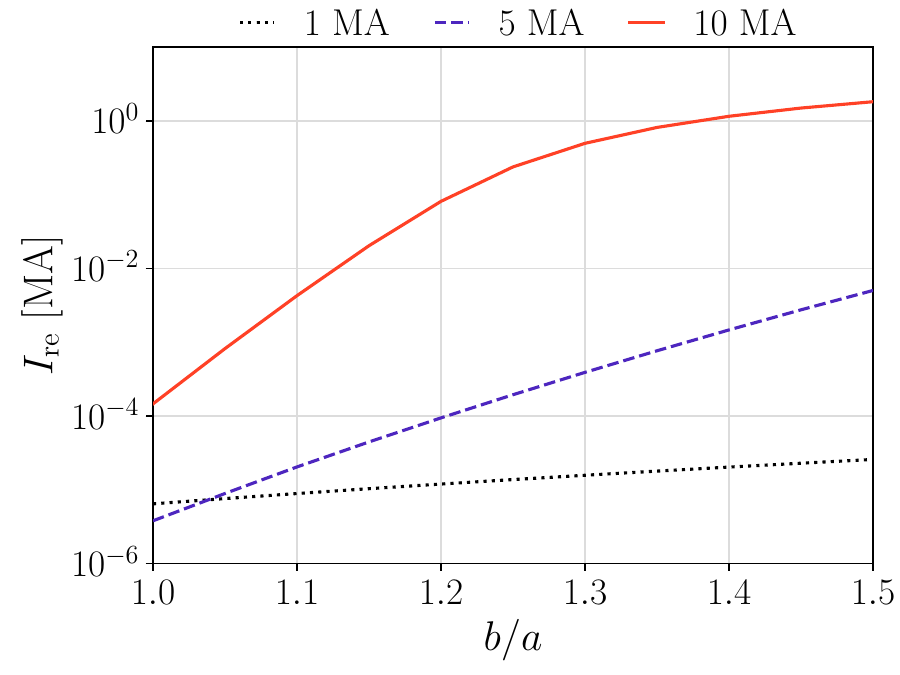}
    \put(-165,146){$I_{\rm p}$:}
    \caption{Generated runaway current as a function of the distance to the toroidally closed conducting structure normalized to the plasma minor radius $b/a$ for an initial plasma current $I_{\rm p}=\SI{1}{MA}$ (black dotted), $I_{\rm p}=\SI{5}{MA}$ (purple dashed), and $I_{\rm p}=\SI{10}{MA}$ (red solid). All simulations correspond to $T_{\rm fin}=\SI{1}{eV}$ and $\tau=\SI{2}{ms}$. }
    \label{fig:scan_fb}
\end{figure}

However, the avalanche multiplication is also sensitively dependent on the distance to the toroidally closed conducting structure, as illustrated in figure \ref{fig:scan_fb}.
The distance to the toroidally closed conducting structure determines the total poloidal magnetic flux of the configuration, which is an important factor in determining the avalanche gain, as detailed in Ref.~\cite{hesslow2019}. 
Figure \ref{fig:scan_fb} presents how the runaway current depends on the distance to the toroidally closed conducting structure $b$, normalized to the minor radius $a$, for a slow temperature collapse ($\tau=\SI{2}{ms}$) with a low final temperature $T_{\rm fin}=\SI{1}{eV}$ and initial plasma current $I_{\rm p}\in\{1,\ 5,\ 10\}\,\si{MA}$. 
Having a long temperature collapse and low final temperature corresponds to having a small runaway seed, but strong avalanche, illustrating how significant the runaway generation can be even if the seed can be limited. 
For a high initial plasma current of $\SI{10}{MA}$, similar to that of a reactor-scale tokamak, the avalanche can be sufficiently strong to generate runaway currents $I_{\rm re}>\SI{100}{kA}$ for $b/a>1.2$. 
However, \SI{10}{MA} is a very high value for a stellarator plasma current, and for $I_{\rm p}\leq\SI{5}{MA}$, the generated runaway current $I_{\rm re}\leq\SI{10}{kA}$ even at $b=1.5a$ for this configuration. 
This further supports the finding that avalanche multiplication will not be a major concern in reactor-scale stellarators with moderate plasma currents. 

\begin{figure}
    \centering
    \includegraphics[width=8.64cm]{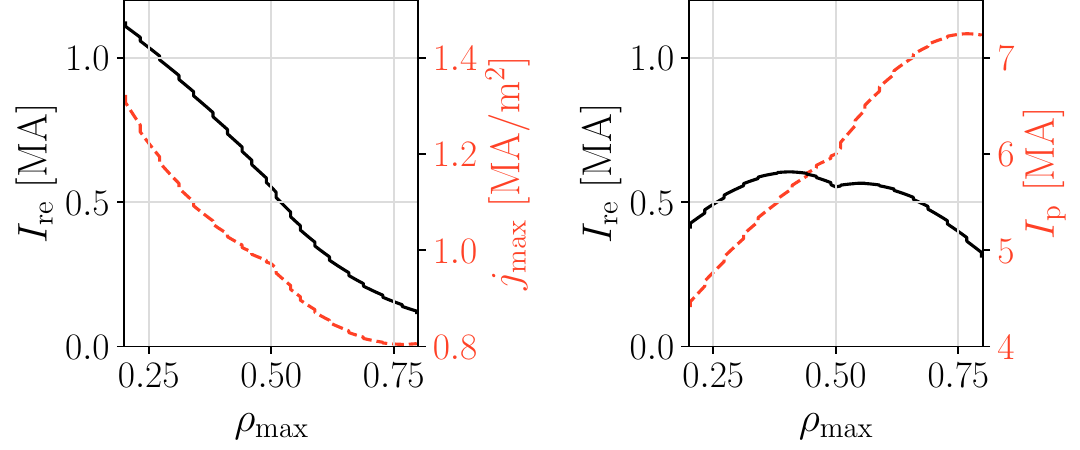}
    \put(-164,94){(a)}
    \put(-87,94){(b)}
    \caption{Generated runaway current (black) as a function of the peak location for different current density profiles (a) with a constant plasma current of \SI{6}{MA}, and (b) with a constant maximum current density of \SI{1}{MA/m^2}. In (a), the maximum current density value is plotted in red with the corresponding red vertical axis on the right. In (b), the total plasma current is plotted in red instead, corresponding to the right vertical axis. }
    \label{fig:profscan}
\end{figure}
\begin{figure}
    \centering
    \includegraphics[width=8.64cm]{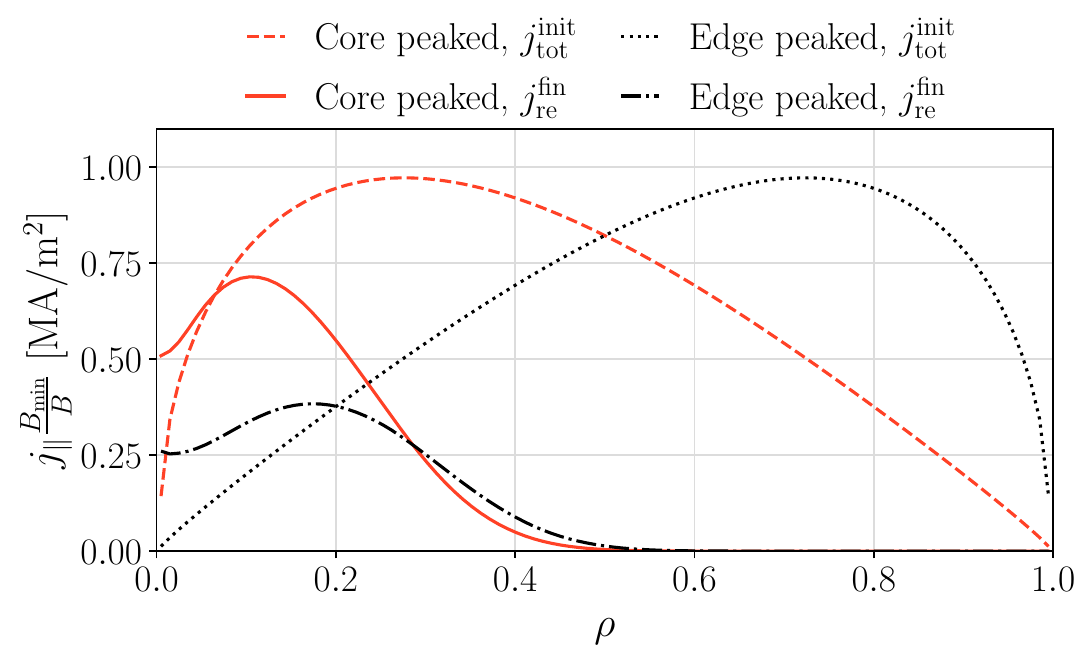}
    \caption{Initial total current density profile and final runaway current density profile for a core-peaked (red) and edge-peaked (black) initial profile.}
    \label{fig:reprofs}
\end{figure}

Aside from the magnitude of the plasma current, the shape of the current density profile affects the runaway dynamics; figure \ref{fig:profscan} demonstrates that a current density profile peaked towards the core is more prone to runaway electron generation.
The sensitivity to the current density profile, keeping the total plasma current fixed (illustrated in figure \ref{fig:initprofBS}a), is presented in \ref{fig:profscan}a. 
With a fixed total plasma current, the runaway current is monotonically increasing as the peak of the initial plasma current gets closer to the core. 
However, as the peak of the initial plasma current gets closer to the core, the maximum current density also increases. 
Both the initial plasma current and current density are important in determining the avalanche multiplication, and it is therefore difficult to distinguish the effect of the location from the effect of the magnitude of the current density. 
The sensitivity to the current density profile, keeping the current density magnitude fixed (illustrated in figure \ref{fig:initprofBS}b), is instead presented in \ref{fig:profscan}b, and now the profile dependence is no longer monotonic. 
For core-peaked profiles, the generated runaway current initially increases with increasing $\rho_{\rm max}$, which describes the location of the peak, which can be explained by the increasing total plasma current. 
However, after $\rho_{\rm max}\approx0.4$, the runaway current decreases with $\rho_{\rm max}$ (except for a flat bump at $\rho_{\rm max}=0.55$) even though the total plasma current is monotonically increasing with $\rho_{\rm max}$. 
Both of the profile sensitivity scans thus demonstrate that more runaway electrons are generated with a core-peaked initial current density, and the reason can be explained by figure \ref{fig:reprofs}. 
Here, the initial plasma current and final runaway current are plotted for a core peaked and an edge-peaked initial profile. 
Regardless of the initial current density profile, the runaway current density is peaked towards the core of the plasma. 
Having a higher initial current density towards the core thus enhances the avalanche multiplication where the runaway electron generation is facilitated. 

\section{\label{sec:con}Discussion and Conclusions}
In this paper, we have presented a pilot study of runaway electrons during temperature collapse in stellarators, which suggests that runaway electron generation can be possible. 
We have explored how the runaway generation is influenced by the temperature collapse time scale, post-collapse temperature, and magnitude of plasma current through numerical simulations using \textsc{Dream}.
Our results predict that there can be significant runaway generation if there is a sufficiently large seed generated by hot-tail generation, which can happen if the thermal quench time is less than \SI{9}{ms}. 
Furthermore, for significant generation two out of three of the following conditions must be fulfilled: (1) large initial plasma current ($I_{\rm p}\gtrsim\SI{5}{MA}$), (2) low post-collapse temperature ($T_{\rm fin}\lesssim\SI{20}{eV}$), and (3) short thermal quench time ($t_{\rm TQ}\lesssim\SI{2.8}{ms}$). 
If the thermal quench time is faster than $\sim\SI{2.8}{ms}$, the hot-tail generation is strong, and if not, strong avalanche generation is required.
Our findings are complementary to, but notably in agreement with, those of Aleynikov \emph{et al.}, who estimate runaway generation to be possible during coil quenching in W7-X, but unlikely during routine operation~\cite{aleynikov2026}. 

We have found the severity of the runaway electron generation is sensitively dependent on the thermal quench time scale, and thermal quench experiments in LHD suggest the thermal quench times in stellarators during radiative collapses could be long enough to make significant runaway generation unlikely. 
Our simulations predict that thermal quench times below \SI{9}{ms} could produce runaway currents over \SI{100}{kA}, depending on the strength of the avalanche generation. 
The thermal quench times of temperature collapses in LHD have been found to be $\gtrsim\SI{50}{ms}$ for injection of $3\times 10^{17}$ tungsten atoms~\cite{bouvain2026}. 
If this is representative of possible unintended radiative collapses in stellarators, it would correspond to the higher end of the $t_{\rm TQ}$ interval we have considered, and runaway generation should not be a concern. 
However, it is possible that the ingress of a larger unintended object, such as a full wall tile, could result in faster temperature collapses that do generate runaway electrons. 
The time scales can possibly also be affected by machine size. 
In Ref.~\cite{bouvain2026}, the radiative collapse was driven by cold fronts propagating inward at constant velocities of 1 -- \SI{5}{m/s}, which suggests size could play a role. 
It has been demonstrated that size does play a role in tokamak disruptions, where longer thermal quenches and slower cold front propagations have been observed for larger minor radii~\cite{bodner2025}. 
This motivates further studies of temperature collapses in stellarators to establish relevant time ranges. 
Ref.~\cite{bouvain2026} also finds that electron cyclotron resonance heating can be used to extend the thermal quench duration, which could offer a possible method for mitigating hot-tail generation in stellarators, should it risk being a problem in future devices.

\begin{figure}
    \centering
    \includegraphics[width=8.64cm]{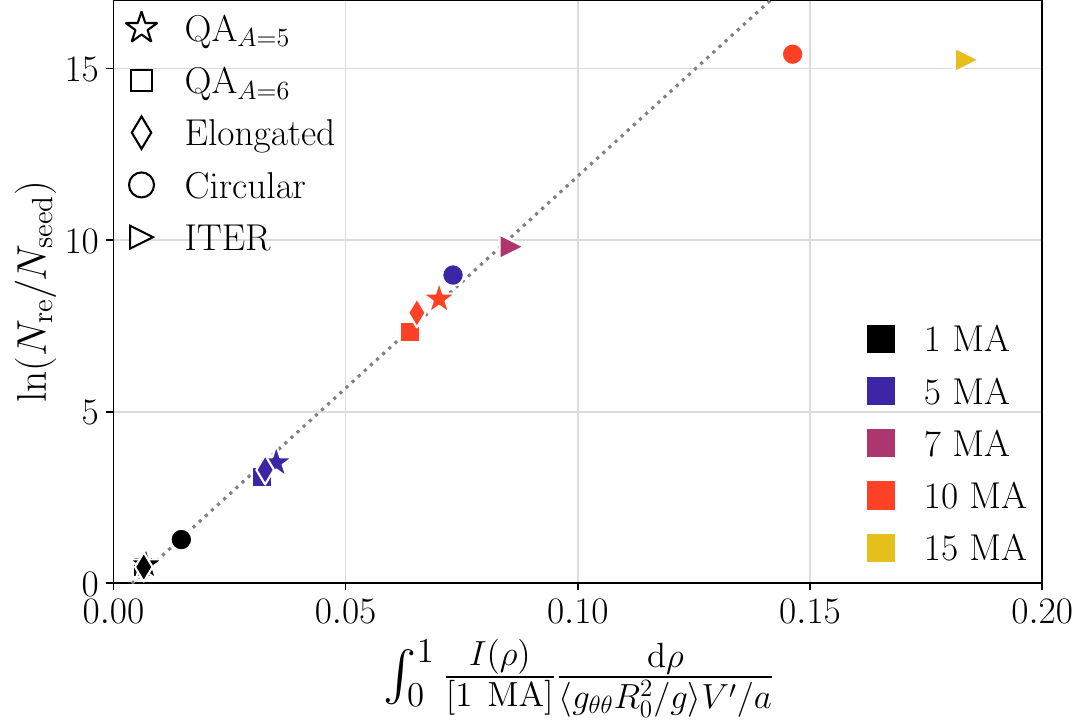}
    \caption{Metric to estimate the logarithm of the avalanche multiplication factor for a given magnetic configuration and plasma current. The metric has been applied to the stellarator configuration used in this work (QA$_{A=5}$), a similar quasi-axisymmetric configuration of aspect ratio $A=6$ (QA$_{A=6}$), and the same circular and elongated plasmas as in figure \ref{fig:Ireshapes}, all for $I_{\rm p}\in\{1,\ 5,\ 10\}\,\si{MA}$. Additionally, an ITER-like configuration is also included for $I_{\rm p}\in\{7,\ 15\}\,\si{MA}$.} 
    \label{fig:metric}
\end{figure}

Interestingly, we conclude both the seed generation rate to be lower and the avalanche multiplication to be weaker in stellarators compared with tokamaks due to the stronger elongation. 
This is because an increased elongation reduces the electric field diffusion, as described in Ref.~\cite{fulop2020}, which has a significant effect on the runaway dynamics. 
Thus elongation can rival plasma current in determining the avalanche gain for a magnetic configuration. 
A metric for estimating the avalanche gain for a given magnetic configuration and plasma current can be found using Ampère's law (see equation \eqref{eq:AmpSt}) to be
\begin{equation}
    \int_0^1\frac{I(\rho)}{[\SI{1}{MA}]}\frac{\dd \rho}{\left\langle g_{\theta\theta}/g\right\rangle V'/a}. 
\end{equation}
Figure \ref{fig:metric} illustrates that this metric correlates with the logarithm of the avalanche multiplication factor $M_{\rm ava}=\ln (N_{\rm re}/N_{\rm seed})$ for both stellarator and tokamak plasmas, where $N_{\rm re}$ and $N_{\rm seed}$ are the runaway and seed particle numbers, respectively. 
The logarithm of the avalanche multiplication factor has been determined from simulations of the different magnetic configurations with wall radius $b=a$, constant temperature $T=\SI{1}{eV}$ and uniform initial seed density of $n_{\rm re}=\SI{5e8}{m^{-3}}$, while the only active runaway generation mechanism is avalanche generation. 
As shown in the figure, the correlation is approximately linear, at least for lower values of the metric. 
For higher values of the metric, the dependence becomes logarithmic rather than linear, suggesting that the avalanche multiplication is limited. 
The limitation comes from the runaway current saturating when it becomes of similar order of magnitude as the initial plasma current. 
The metric represents how severe the avalanche multiplication could be in a given device, and could be included in magnetic configuration optimization to account for runaway electron avoidance. 
As for the magnetic configurations  considered, figure \ref{fig:metric} illustrates that the stellarator plasmas, and the elongated tokamak plasma illustrated in figure \ref{fig:Ireshapes}d, are significantly less prone to avalanche generation than both a circular and an ITER-like plasma. 
The strongly shaped plasmas have a lower avalanche multiplication factor for $I_{\rm p}=\SI{10}{MA}$ than the circular plasma has for $\SI{5}{MA}$ and the ITER-like plasma for $I_{\rm p}=\SI{7}{MA}$. 
In conclusion, the combination of low plasma currents and strong elongation limits the avalanche multiplication of stellarator configurations, and in turn the risk of significant runaway generation. 

The dependence of the avalanche multiplication on the plasma configuration through elongation demonstrates the limitation of only using one stellarator configuration for this study; the strength of the avalanche multiplication, and in turn the risk of runaway electron generation, is intrinsically configuration dependent. 
Thus, to further explore the intricacies of runaway electrons in stellarators, a wider set of stellarator configurations should be considered. 
In particular, it would be of interest to apply the same considerations to reactor-relevant stellarators with consistent configurations and radial profiles. 
Yet, since the elongation mainly determines the scaling of the avalanche multiplication factor, we believe that the general trends predicted in this paper are representative for stellarators in general. 

Using a prescribed temperature evolution presents two other important limitations to consider: the presence of impurities will determine the temperature evolution, which is crucial for modelling the runaway electron dynamics, but it will not only affect the runaway generation through the temperature evolution. 
The presence of impurities that are partially ionized will also affect the runaway generation directly, both through the critical electric field evolution and because the avalanche generation is proportional to the total density of electrons. 
A more robust and self-consistent treatment of the evolution of the temperature and plasma composition is possible in \textsc{Dream}, and presents an interesting avenue for future studies. 

\begin{acknowledgments}
The authors would like to extend their gratitude to Rory Conlin, Tünde Fülöp, Per Helander, Matthias Hölzl, Mathias Hoppe, Jens von der Linden, Sarah Newton and István Pusztai for fruitful discussions. 
The authors are also grateful to Stefan Buller for providing a suitable magnetic configuration for the simulations. 

This work was supported by the Swedish Research Council (Dnr. 2022-02862), and by the Knut and Alice Wallenberg foundation (Dnr. 2022.0087 and 2023.0249). This work has been carried out within the framework of the EUROfusion Consortium, funded by the European Union via the Euratom Research and Training Programme (Grant Agreement No 101052200 — EUROfusion). 
M.L.~was supported by the US Department of Energy under contract DE-FG02-93ER54197.

Views and opinions expressed are however those of the author only and do not necessarily reflect those of the European Union or the European Commission. Neither the European Union nor the European Commission can be held responsible for them.

\end{acknowledgments}

\section*{DATA AVAILABILITY}
The data that support the findings of this study were generated by numerical simulations. 
The source code \textsc{DREAM} used to generate the simulations is publicly available~\cite{dream}.
To run the simulations, the DESC equilibrium solver Python module is needed and is publicly available~\cite{dudt2020}. 
The magnetic configuration needed for running the stellarator simulations is publicly available~[34]. Simulation codes and the data that support the findings of this article (Figs.~4–11) are publicly available~[33].

\appendix

\bibliography{ref}

\end{document}